\documentclass[]{spie}  
\pdfoutput=1
\usepackage[]{graphicx}
\usepackage[utf8]{inputenc}
\usepackage{import}
\usepackage{url}

\title{An Application of CNNs to Time Sequenced One Dimensional Data in Radiation Detection} 

\author{Eric T. Moore\supit{a}, William P. Ford\supit{a}, Emma J. Hague\supit{a}, and Johanna Turk\supit{b}
\skiplinehalf
\supit{a}Remote Sensing Laboratory, Joint Base Andrews \\
\supit{b}Barnstorm Research, Boston, MA
}

\authorinfo{Further author information: (Send correspondence to E.T. Moore)\\ E.T. Moore: E-mail: mooreet@nv.doe.gov, Telephone: 1 703 853 7994}

\begin{document} 
  \maketitle 

\begin{abstract}
A Convolutional Neural Network architecture was used to classify various isotopes of time-sequenced gamma-ray spectra, a typical output of a radiation detection system of a type commonly fielded for security or environmental measurement purposes. A two-dimensional surface (waterfall plot) in time-energy space is interpreted as a monochromatic image and standard image-based CNN techniques are applied. This allows for the time-sequenced aspects of features in the data to be discovered by the network, as opposed to standard algorithms which arbitrarily time bin the data to satisfy the intuition of a human spectroscopist. The CNN architecture and results are presented along with a comparison to conventional techniques. The results of this novel application of image processing techniques to radiation data will be presented along with a comparison to more conventional adaptive methods \cite{Ford18}.
\end{abstract}



\section{INTRODUCTION} \label{sec:intro}
This research is motivated by the lack of robust automated identification algorithms for gamma ray spectroscopy, specifically when the spectra's statistics are low, including potentially low signal$/$noise ratios.  In radiological search operations, source encounters are typically short, usually in the range of $0.5-2.5$ seconds, resulting in low-statistics.  Most tools, both automated and those meant to be employed by a human spectroscopist, are designed to handle spectral data of much longer duration, resulting in high-statistics data.  Traditionally search has focused on “detect”, “localize”, “identify” as separate steps in the process.  However, today many more detection instruments have multi-channel analyzers (MCAs) imbedded, which has resulted in an exponential growth in the availability of spectral information.  Source encounters can be operationally difficult or dangerous to duplicate, so analysis based on available data, rather than a more desirable longer measurement, is critical. This represents a significant change in the concept of operations, to which ‘reach$-$back’ entities have not yet adjusted.  In this modern search environment it is critical, from a technical, operational, and political perspective that short source encounter data can be adjudicated with as much confidence as is reasonably achievable given the limitations of statistics.  And it is quite clear that current methodologies do not begin to approach true precision limits.

A number of adaptive algorithms have been employed over a period of several years for isotopic identification of gamma spectra, but these have been focused primarily on 1-D spectra and they classify these spectra one dimensionally, just as any human spectroscopist would. In this work, we refocus on the data as they come into the instrument, in a time sequence, moving away from the inherently human-centric bias of prior data formatting approaches.  Some of the previous efforts have been focused on very specific aspects of the problem, like class imbalance \cite{Sharma12}.  While other efforts were more general isotopic identification \cite{Portnoy04, Ford18, Jones14}.  Most of these efforts have suffered from a dearth of data in the domain space of interest, such as threat materials; this fact will be addressed later in the paper.  Currently, large data sets are available in the domains of: background, medical, and industrial sources.  The large data sets now available for certain types of sources are what is motivating the strong push toward machine learning approaches.

\section{APPROACH} \label{sec:approach}
Convolutional neural networks (CNNs) are frequently used in the processing of image data \cite{Lecun98, Torrey09}.  CNNs are ideally suited for this type of data due in large part to their tendency to reduce the dimensionality of the data in a natural way, which leads to a more computationally realistic problem \cite{alex}.  CNNs have spawned rapid growth in the application of ever deeper neural architectures \cite{Simonyan15}, and in propagating trained architectures and libraries to facilitate the development of new architectures \cite{caret,scikit,tensorflow,keras}.

\begin{figure}[ht]
	\begin{center}
		\begin{tabular}{c}
			\includegraphics[height=9.5cm]{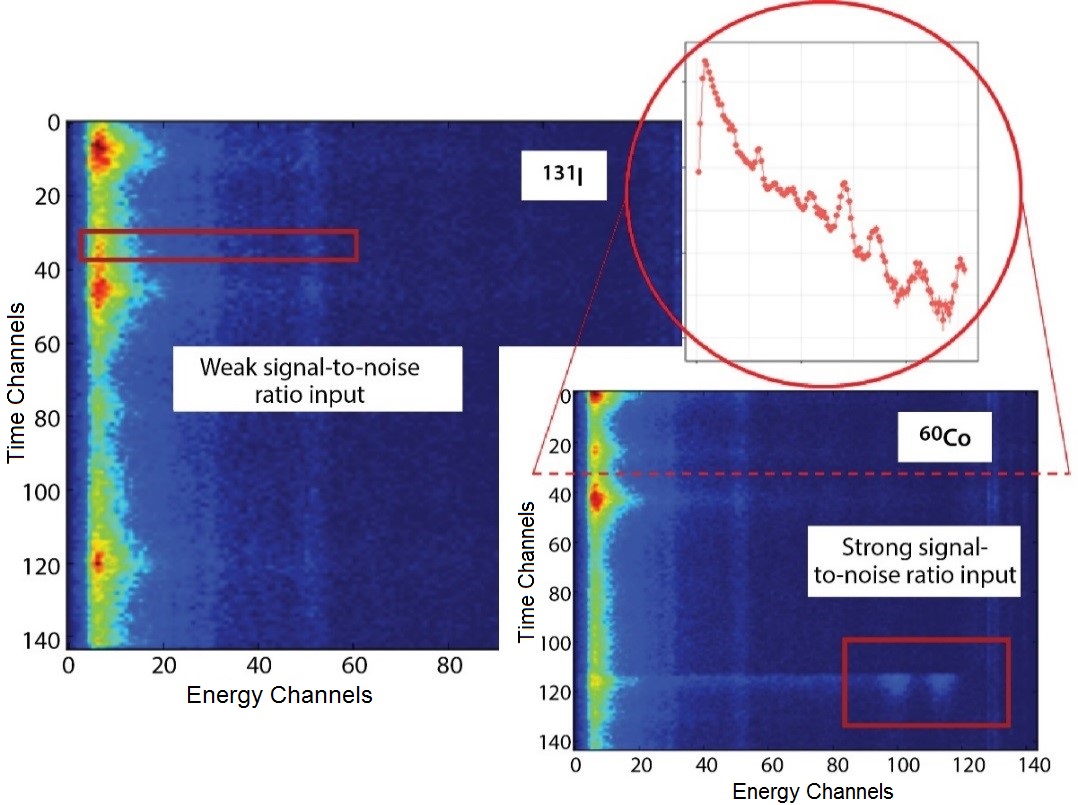}
		\end{tabular}
		
	\end{center}
	\caption[example] 
	{ \label{fig:waterfall} 
		Waterfall plots (false color of monochromatic images, $144 \times  144$ pixels) illustrate the nature of the input used for training and testing the network. The signal-to-noise ratio may be strong or weak (the red boxes are meant to illustrate this, and do not correspond to an object detection box). The three-second time slice from the run (inset red circle) illustrates how the data are formatted.}
\end{figure}

The significant research in the use of CNNs for the processing of image data has motivated us to exploit these techniques in our (1-D plus time) domain space, allowing us to transform time-sequenced 1-D data into 2-D monochromatic image data (Fig.~\ref{fig:waterfall}).  Each input image is $144 \times  144$ pixels (temporal channels versus energy channels), and the time axis is “stretched” to fit this number of bins; thus, some runs may have one-quarter-second binning and some multiple-second binning depending on the varying lengths of individual runs. A max-one normalization is the only other pre-processing performed on the images.  This approach should allow the application of various 2-D algorithms to discover the time dependence of source encounters in the course of the training process.

Visualizing data as false-color waterfall images has been relatively common in radiation detection operations.  The y-axis is time and the x-axis is energy of incoming photons.  A cross section along $x$ at a fixed time $y$ yields a gamma ray energy spectrum (histogram) as shown in the inset Fig.~\ref{fig:waterfall}.  The color on the waterfall image represents height or intensity (counts) at a given time and energy coordinate.  The image sizes are shown as an example, and multiple input image sizes were tested with the various architectures employed.

The intent of this work was not to exhaust all the possible options for processing data in this way, and only a few general architecture types have thus far been employed.  The first category of CNNs employed used multiple stacked convolutional and pooling layers without skip-connections \cite{Lecun98, alex}; similar deeper networks with skip-connection where also tested \cite{resnet}.  And finally a group of ‘wider’ networks using inception modules was employed \cite{inception}.  Although no great variation in performance was found in the various types of architectures employed it was not the intent of this work to optimize a given architectural solution, but rather to highlight the general method of dealing with this type of dataset; in any case, it is unlikely that the dataset employed was large and diverse enough to be used for optimizing performance of the networks, but we have confidence that this can be done with other datasets that are now becoming available to the community \cite{novarray}.  Additionally, a 1-D approach will be shown for comparison only.

\section{DATASET} \label{sec:data}
\begin{figure}[h!]
	\begin{center}
		\begin{tabular}{c}
			\includegraphics[height=6.5cm]{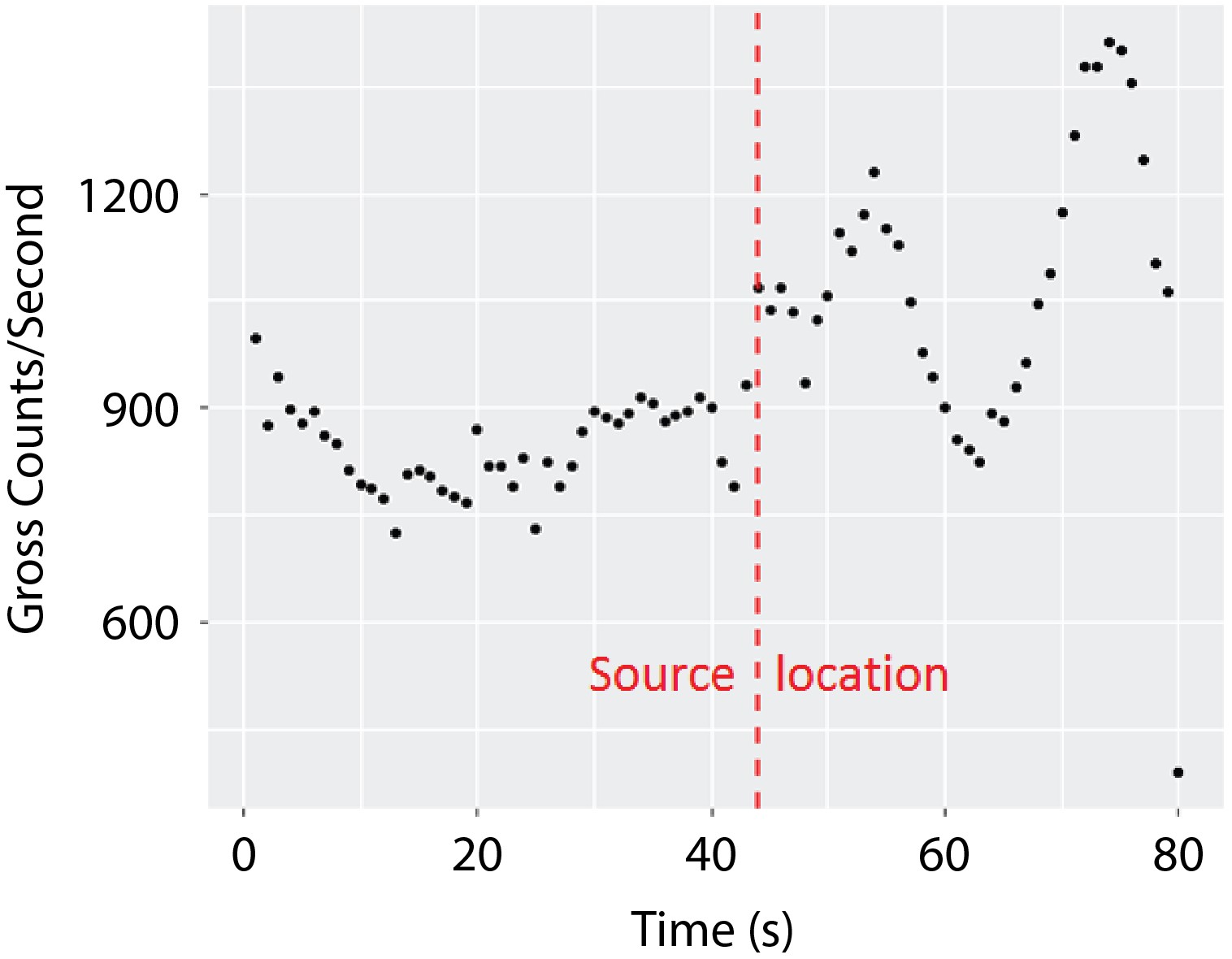}
		\end{tabular}
	\end{center}
	\caption[example] 
	{ \label{fig:ggc} 
		The number of photons detected in the measured energy range per second for one run. It can be seen that any anomaly detection must be isotopically based.}
\end{figure}

The data we used came from a data competition (Urban Radiological Search Data Competition) run by Lawrence Berkley National Laboratory (LBNL) for the DOE Office of Defense Nuclear Nonproliferation Research and Development \cite{URSC}. The simulated spectral data were supplied in list mode format (i.e., individual photon energy and arrival time). The data were then binned into time and energy spectra (Fig.~\ref{fig:waterfall} inset) as described above. Variation in background radiation typical of mobile detector data is seen in the data set, and therefore peaks in gross count rate do not necessarily correspond with source location (Fig.~\ref{fig:ggc}).

The spectra used were modeled in a simulated city (background), with sources placed at random locations throughout (gamma ray readings only).  There are six source classes, each including sub-classes having varying activity and shielding located within the simulated city, and one class representing background only, for a total of seven classes. The six simulated source classes are highly enriched uranium (HEU), weapons grade plutonium (WGPu), $^{131}$I, $^{60}$Co, $^{99m}$Tc, and a combination of HEU and $^{99m}$Tc.  The training set contains 9800 samples (`runs'): 5000 background and 800 of each of the source classes.  In the runs containing a source, only one such source will be present in any given run, and runs vary substantially in duration. The background varies throughout the city, but is not altered from one run to the next; the duration of the run varies, and the speed of the moving detector also varies.

There was a concern that the similarity of the background from one run to the next might introduce a bias. Therefore, two separate input image sets were constructed: one with the full background time period (which varied between runs), and one with 60 seconds of background enveloping the source. These two data sets were fed through two differently constructed networks, essentially giving separate analyses on data formatted differently along the time axis. The time binning in the first dataset (the y-axis in Fig.~\ref{fig:waterfall})) is of varying bin width, simply the total time of the run divided into $144$ bins. To compensate for the potential bias of identical background for different runs, we separated out thirty-second time intervals from each run to construct the second dataset; the selected time interval includes the source, but it falls into a randomly chosen time window within the newly defined run, and this new waterfall is then used as the input image. Excluding the 2 seconds at the start and end of each 30-second sampling, the source location is set via a unitary sampling along the y-axis (time), so there would be no preferred location. The second data set includes different background regions across the whole space of possible runs.  Because both the aforementioned data constructions yielded similar results, we show only one result; however, it is worth noting that the network is robust against varying time binning or stretching/contracting of the image along the {\it temporal} axis.

\section{ARCHITECTURE} \label{sec:arch}
We used multiple machine learning algorithms, as discussed above, to classify the dataset \cite{URSC}.  All of these involve the dataset being preprocessed as described in the preceding two sections, although the exact image sizes may vary.  The architectures we investigated were chosen because of their proven success with other image classification problems \cite{Lecun98, alex, resnet, inception}. In addition, we are currently working to apply transfer learning techniques to radiological data but this work is addressed elsewhere \cite{Moore19}.

\subsection{Base Convolutional Structure} \label{ssec:vggnet}

The first architecture begins with ten convolutional layers, and every other convolutional layer is followed by a max pooling layer. The output of the convolutional and pooling layers is then flattened to create a dense layer that feeds into three fully connected layers.  A rectified linear unit (ReLU) non-linearity is applied to both the convolutional and dense layers; it still needs to be demonstrated that ‘dying’ nodes are not a problem. The output of the three dense (fully connected) layers is fed into a seven-class output layer with softmax ($ \sigma$), which is a typical normalization method for categorical distributions,
\[ \sigma (z)_j = \frac{e^{z_j} }{\sum_{k=0}^{K} e^{z_k} }, \hspace{0.75 cm} for \hspace{0.25 cm} j=1,\mathellipsis ,K ,\] \\
where $\sigma$ is given by the standard exponential function at each coordinate, divided by the sum of the exponential function applied to each coordinate, so the output coordinates sum to unity. This arrangement gives a total of 14 layers with trainable weights, and the total number of trainable parameters (weights) in the network is approximately 8.2 million.

The architecture employed was heavily influenced by the AlexNet and VGGnet architectures \cite{Lecun98,alex, Simonyan15}.  Most of the results shown in this work will be from this type of architecture; but a number of different architectures were tried, loosely based on other successful image classification architectures, and further architectures are being tested \cite{resnet,inception}.  The final selection of an architecture is very much driven by the size and representativeness of the dataset.

During training the network converges relatively quickly, in approximately 25 to 50 epochs, using a batch size of 75 on a training set of 5400 waterfall images. The top half of Fig.~\ref{fig:vgg} shows the convergence within 50 training epochs. Convergence is considered to occur as the cost function is minimized; the cost is given by the cross entropy $E$,
\[ E = \frac{1}{N} \sum_{n=0}^{N} y_n \log(p_n)  +  (1-y_n) \log(1-p_n), \]
where $p$ refers to the predicted value for an event and $y$ to the label (true) class for each event. Half of the training runs are background only , and the other half are equally divided over the remaining six source classes. Due to the abundance of background events, too few training events or too complex an architecture can lead to `trapping' all the events in the background class; however, it should be remembered that in real data, there are a number of background types, so this situation is not easily fixed by reducing the proportion of background. Alternatively, we could try to use a separate detection algorithm to pick out `hits' first; issues related to class inbalance are beyond the scope of this paper.

The training algorithm, as implemented, can be run in a reasonable time on a single GPU; running without a GPU, even on multiple cores, has been too slow for the tuning of the hyper-parameters, which has not yet been completed but would be necessary if the objective where an operational algorithm.

\subsection{Residual Architecture} \label{ssec:resnet}

\begin{figure}[ht]
	\begin{center}
		\begin{tabular}{c}
			\includegraphics[height=8cm]{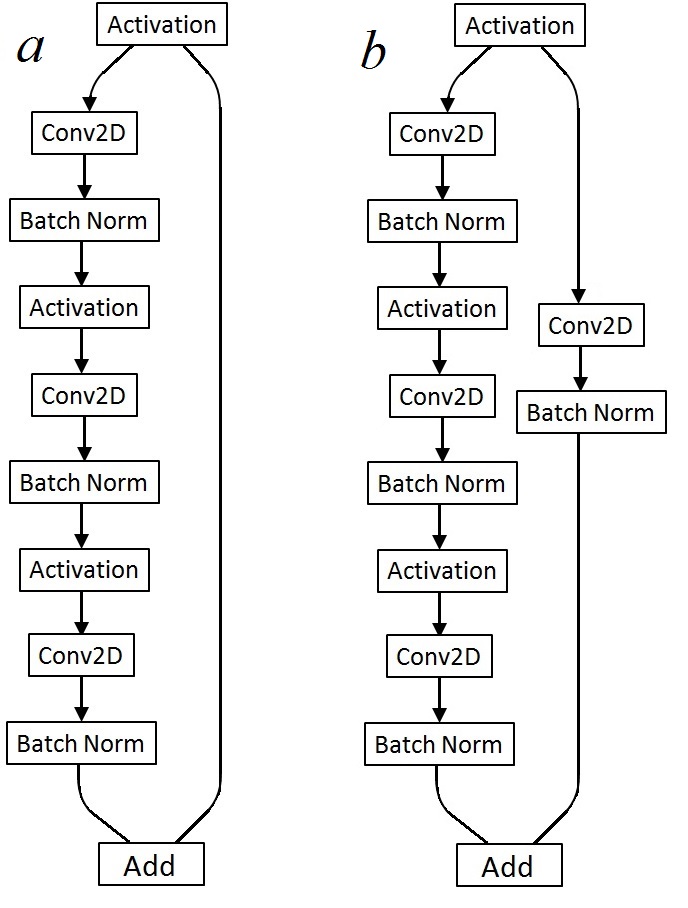}
		\end{tabular}
	\end{center}
	\caption[example] 
	{ \label{fig:resnet_blocks} 
		{\bf (a)} Diagram of a identity block in the residual network structure.  {\bf (b)} Diagram of a convolutional block in the residual network structure.}
\end{figure}

The residual architecture, with skip connections, is based on the ResNet architecture \cite{resnet}.  The implementation that we have used is $50$~layers and has approximately $24$~million trainable parameters and we drew heavily on a similar implementation used built for classification of the {\it CIFAR-10} dataset \cite{Dwivedi19}.  The overall architecture is shown in Table~\ref{table:total_resnet}, which consists of a number of stages.  Each stage in Table~\ref{table:total_resnet} represents a number of blocks within the overall network.  In turn each block is a consistent sequence of layers for the given type of block.  The structure of a convolutional block is shown in Fig.~\ref{fig:resnet_blocks};  the structure of the identity block is not shown but is the same except the shortcut route has no convolution and normalization, it is just a bypass \cite{resnet}.

\begin{table}[ht]
	\caption{Overall Structure of Network with Residual Architecture} 
	\centering 
	\begin{tabular}{c c c c} 
		\hline\hline 
		Stage & Total~\# of Blocks & \#~Convolutional Blocks & \#~Identity Blocks \\ [0.5ex] 
		\hline 
		Pre-stage & \multicolumn{3}{c}{{\it No blocks:} input $\rightarrow$ Conv2D $\rightarrow$ Norm $\rightarrow$ Max Pooling $\rightarrow$ stage~1} \\ 
		1 & 3 & 1 & 2 \\
		2 & 4 & 1 & 3 \\
		3 & 6 & 1 & 5 \\
		4 & 3 & 1 & 2 \\ 
		Post-stage & \multicolumn{3}{c}{{\it No blocks:} stage 4 $\rightarrow$ Avg Pooling $\rightarrow$ flatten $\rightarrow$ Dense (output layer) } \\ [1ex] 
		\hline 
	\end{tabular}
	\label{table:total_resnet} 
	
	\vspace{0.5cm}
\end{table}

\subsection{Inception Architecture} \label{ssec:inception}

\begin{figure}[ht]
	\begin{center}
		\begin{tabular}{c}
			\includegraphics[height=5.5cm]{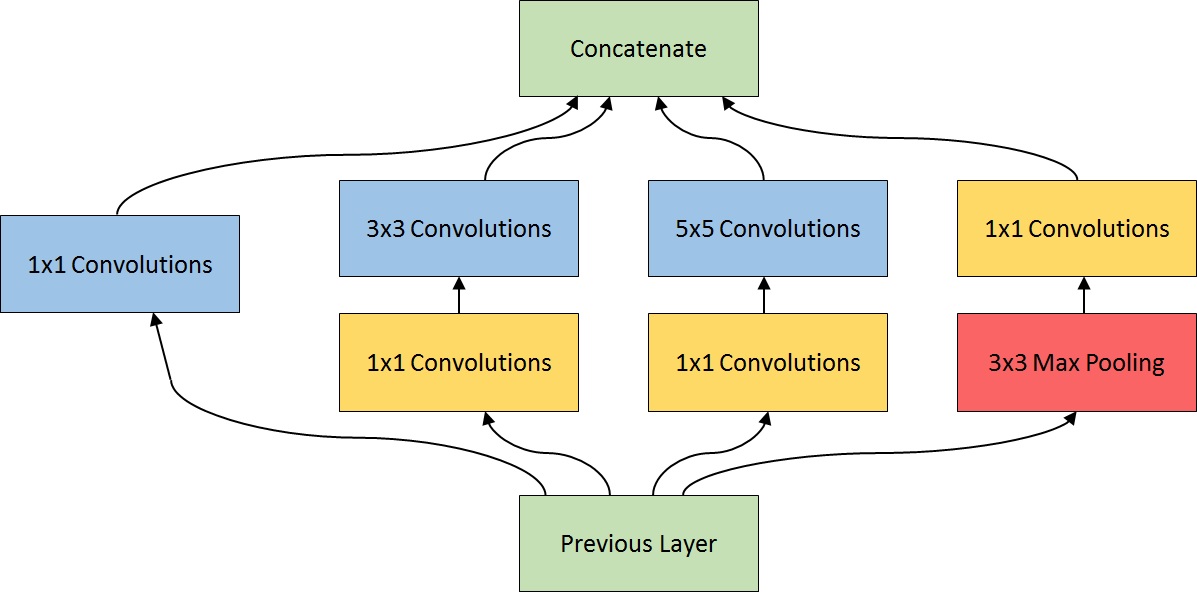}
		\end{tabular}
	\end{center}
	\caption[example] 
	{ \label{fig:inception_module} 
		Inception module with dimensional reduction \cite{inception}.}
\end{figure}

A $64 \times 64$ image set is generated for use in this and the residual architecture in Section~\ref{ssec:resnet}.  The inception module architecture has proven valuable in other image classification problems, and detailed descriptions are readily available elsewhere \cite{inception}; however a diagram of the inception module is included in Figure \ref{fig:inception_module} to emphasize the differing nature of this type of structure \ref{fig:inception_module}; where multiple size convolutions are performed simultaneously at each stage, thus creating a {\it wider} rather than {\it deeper} structure.

\section{RESULTS} \label{sec:results}
The focus of this paper is on the technique of utilizing image processing techniques in the domain of gamma ray spectroscopy, where the $x$-axis of the image represents energy and the $y$-axis represents time. The primary results are focused on the performance of CNN without skip-connections, and on classification techniques for gamma-ray spectra formatted as monochromatic images; however we also present one of the one-dimensional analyses that have been done for comparison, and a limited set of results from other CNN architectures.

\begin{figure}[ht]
	\begin{center}
		\begin{tabular}{c}
			\includegraphics[height=8.5cm]{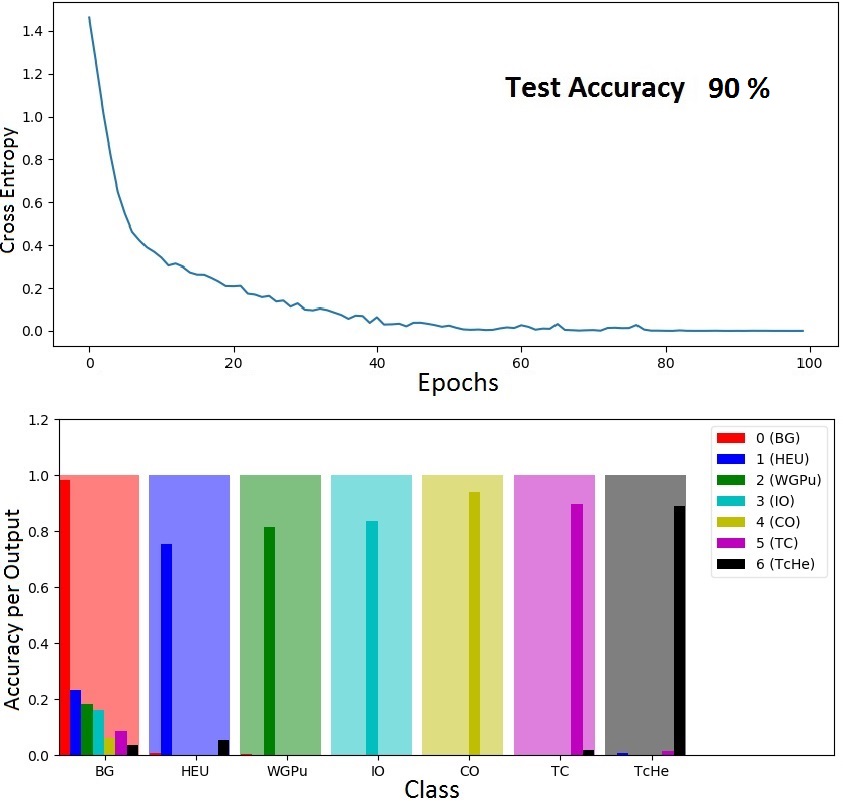}
		\end{tabular}
	\end{center}
	\caption[example] 
	{ \label{fig:vgg} 
		The colored bands are the Base CNN-predicted classes, and the colored bars represent the true classes; if bar and block colors match, the classification is correct. The main area of failure is source events being classified as background (false negatives). The classes are: background, HEU, WGPu, $^{131}$I, $^{60}$Co, $^{99m}$Tc, and HEU combined with $^{99m}$Tc.}
\end{figure} 

\subsection{Metric Importance—Accuracy vs. Recall} \label{ssec:metric}

Although we primarily report accuracy in this paper, it should be noted that we would, in an operational setting, have a very low tolerance for false negatives in the target domain; therefore, it might be more appropriate to maximize recall. Although false negatives are a significant concern for the problem domain, this paper is not focused on the end solution but on the initial application of the methodology; that being said, we have begun to investigate the false negative rates in particular, and the current state of those conclusions is shown below.

\subsection{Accuracy—Base CNN} \label{ssec:accuracy}

The overall accuracy of the Base CNN was $90 \% $, the algorithm performed worse on the threat classes than medical/industrial classes (Fig.~\ref{fig:vgg}).  The network’s ability to distinguish one source from another is good; the greatest shortfall is that source spectra are too often classified as background. Ford \cite{Ford18} and Zeiler \cite{Zeiler14} present more exhaustive treatments of 1-D methods, and examples of feature maps.

The output layer of the network yields a softmax that can be characterized as a kind of confidence level. The default is to choose as the prediction the class that has the highest associated softmax value. Approximately a third of the false negatives choose the true source as their second choice, and this may indicate that more training on a larger dataset might allow for improvement on a signal-to-noise ratio limit \cite{novarray}.

\subsection{1-D Comparison—Random Forest (RF)} \label{ssec:rf}

In addition to Base CNN, we applied multiple 1-D algorithms to the dataset \cite{URSC}.  A random forest (RF) algorithm \cite{caret, scikit} was used over the 1-D spectra over the whole ‘run’ and then clustered by like classifications in semi-sequential events.  The dataset was binned in energy and time, but the input ($x_i$) is a 1-D energy vector, individual spectrum (inset Fig.~\ref{fig:waterfall}), so that a single waterfall will be many individual spectra where each must individually be classified; if there are enough of one class within a prescribed time window, the run as a whole is assigned into that particular class. Only one class per run is allowed, but the clustering requirement prevents multiple assignment in almost all cases.  This and other 1-D methods are more fully described in Ford \cite{sorma18}.{
	\begin{figure}[ht]
		\begin{center}
			\begin{tabular}{c}
				\includegraphics[height=4.5cm]{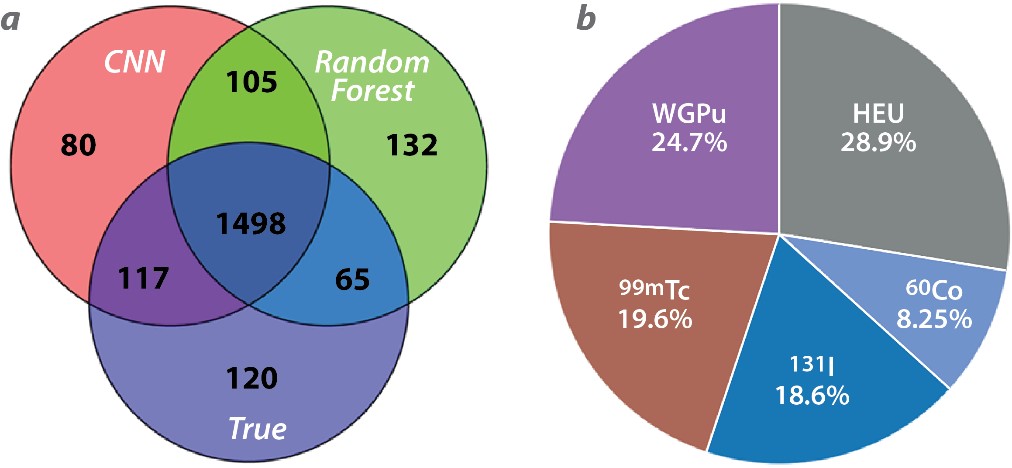}
			\end{tabular}
		\end{center}
		\caption[example] 
		{ \label{fig:venn} 
			(a) The Venn diagram shows all event overlaps. (b) A pie chart of the 87 events that both the CNN and RF incorrectly classified as background (double false negatives).}
	\end{figure} 
}  

The Venn diagram (Fig.~\ref{fig:venn}a) shows all event overlaps; the red circle, for example, shows the Base CNN properly classified 1615 $(1498 + 117)$ events, and misclassified 185 $(105 + 80)$ events. Both the Base CNN and random forest algorithms misclassified 120 events. The RF correctly classified 65 events that the Base CNN misclassified, and the Base CNN correctly classified 117 events that the Random Forest misclassified. The results suggest that combining the two methods might improve overall performance.

\begin{figure}[ht]
	\begin{center}
		\begin{tabular}{c}
			\includegraphics[height=7cm]{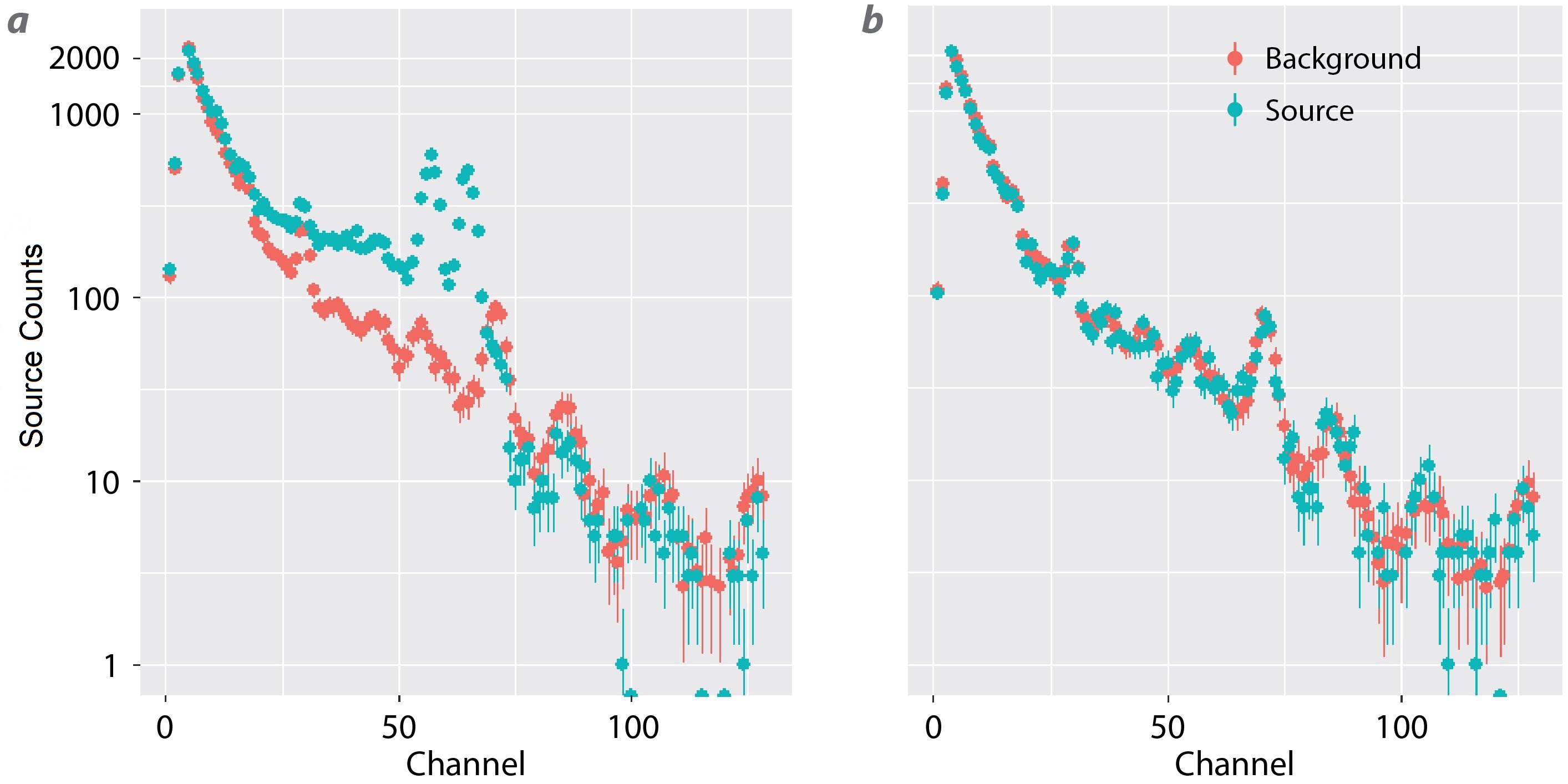}
		\end{tabular}
	\end{center}
	\caption[example] 
	{ \label{fig:spectra} 
		{\bf (a)} A spectrum correctly classified by both the CNN and RF, and {\bf (b)} a spectrum ‘misclassified’  by both the CNN and RF as background, but was labled $^{60}$Co; at some signal:noise ratio the labeling is of questionable validity. The asymptotic background, shown for reference, is scaled to the maximum bin height of the much shorter signal spectrum.}
	\vspace{0.75cm}
\end{figure} 
\begin{figure}[ht]
	\begin{center}
		\begin{tabular}{c}
			\includegraphics[height=8.5cm]{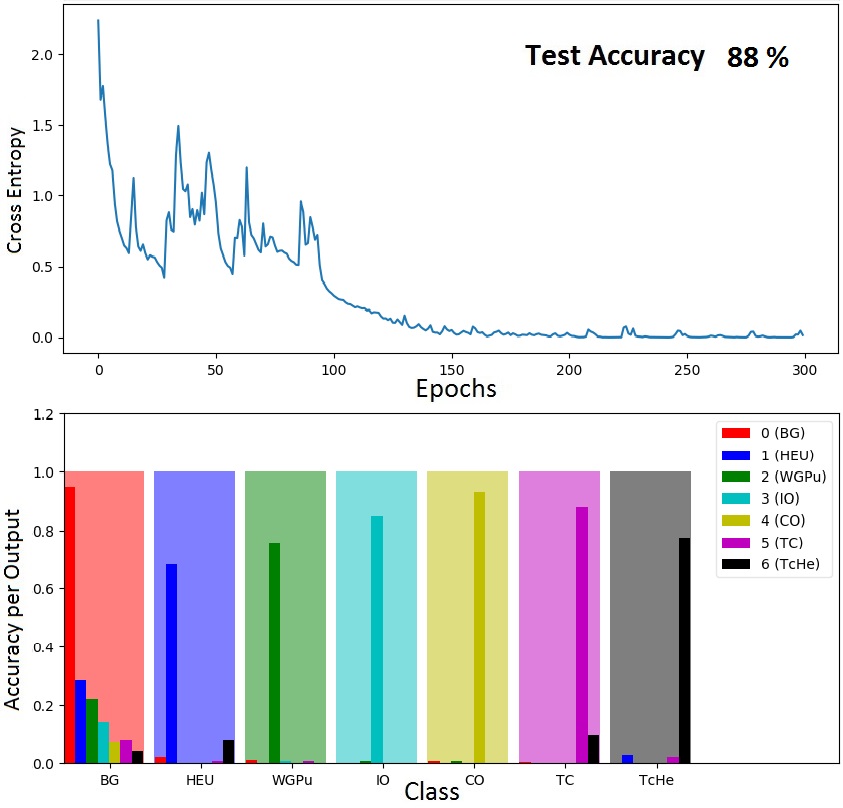}
		\end{tabular}
	\end{center}
	\caption[example] 
	{ \label{fig:resNet} 
		As in Fig.~\ref{fig:vgg} the colored bands are the predicted classes for the network which includes skip connections \cite{resnet}, and the colored bars represent the true classes; if bar and block colors match, the classification is correct.}
\end{figure} 

The total number of events misclassified by the CNN was 185 of 1800 in the testing set.  The total number of events misclassified by the RF was 237 of 1800 on the same testing set.  As shown in the Venn diagram in Fig.~\ref{fig:venn}, the number of events properly classified by both methods was 1498. The total number of events properly classified by either the Base CNN or the RF or both was 1680~$(1498 + 117 + 65)$. The total misclassifications by both the Base CNN and RF, the intersection of error, is 120 events, and of these 105 are misclassified in the same way by both algorithms. Of the 105 events misclassified in the same way by both algorithms 87 of them were false negatives by both algorithms (predicted background when a source was present). It is worth examining these double false negatives in detail to see if classification is realistically possible by any method. The signal-to-noise ratio of the dataset is not known \textit{a priori}; many of the double false negatives may be the result of an absence of signal. Although this question is still being studied, of the cases that have so far been examined, this is the case (Fig.~\ref{fig:spectra}b). In the example shown in this figure, no deviation from background can be seen by a human spectroscopist despite being labeled as $^{60}$Co.

\begin{figure}[ht]
	\begin{center}
		\begin{tabular}{c}
			\includegraphics[height=8.5cm]{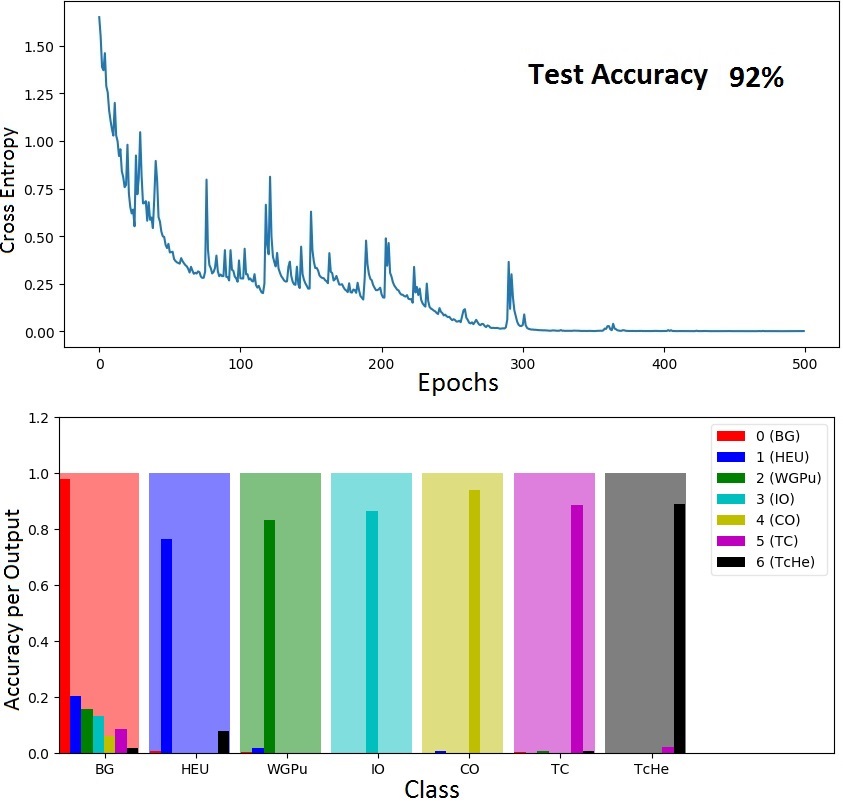}
		\end{tabular}
	\end{center}
	\caption[example] 
	{ \label{fig:inception} 
		As in Fig.~\ref{fig:vgg} the colored bands are the predicted classes in the architecture which includes inception modules \cite{inception}, and the colored bars represent the true classes; if bar and block colors match, the classification is correct.}
\end{figure}

Result for the residual and inception type architectures described in section Sec.~\ref{ssec:resnet}~\&~\ref{ssec:inception} are shown in Figs.~\ref{fig:resNet}~\&~\ref{fig:inception} respectively. It can be seen that although these networks were slower to converge, and somewhat less stable in the training process the results where similar.  Variation in accuracy should be considered in the context that the architectures where not optimized in any meaningful sense; this lack of optimization having to do with the limitations of available data.

As in the comparison above of the basic CNN structure and the RF, we paid particular attention to events which where classified by all four algorithms to be background when they where labeled as source.  In these $44$ of the total $1320$ test events the spectra where examined by a human spectroscopist.  In none of these events could a significant deviation from background be identified; these are then considered to be close the bounding minimum level of reasonable signal given the background rate.

Lastly it should be pointed out that all four methods could in principle be ensembled together to improve overall performance.  However, given the limitations on the dataset, including number of events, it was not worthwhile to explore. Ensembling will be explored in future work with a larger dataset. 

\section{CONCLUSION} \label{sec:conclusion}
Although there is a non-neglible body of work demonstrating the utility of machine learning techniques in gamma ray spectroscopy \cite{Portnoy04,Sharma12,Jones14,Ford18}, these methods are just starting to reach their full potential in this area due to the expanding availability of large datasets.  Gamma ray spectroscopy presents problems in terms of the binning and formatting of the data not generally seen in image processing.  Despite the differences in gamma data we have had good results in utilizing the techniques common in image processing; we hope that this work will help to highlight the importance of optimization of data collection and formatting schema that allow for expanded algorithmic frameworks to be brought to bear on the question of automated gamma spectroscopy.  And we are utilizing these techniques to make it easier to use existing transfer learning methods \cite{Pan10, Yosinski14, Ruder17} for classification of spectra in threat domains where large datasets are difficult or impossible to get.  The transfer learning techniques will be presented in a separate publication \cite{Moore19}.

\acknowledgements
We would like to acknowledge the Department of Energy’s support through the Office of Defense Nuclear Nonproliferation Research and Development NA-22 and the contributors to the \cite{URSC} (`Urban Radiological Search Data Competition') for the dataset.  And we would also like to acknowledge the support of the Site Directed Research and Development (SDRD) program for the National Nuclear Security Site (NNSS) \cite{1DSDRD,CNNSDRD}.

This manuscript has been authored in part by Mission Support and Test Services, LLC, under Contract~No. DE-NA0003624 with the U.S. Department of Energy and supported by the Site-Directed Research and Development Program, National Nuclear Security Administration, USDOE Office of Defense Nuclear Nonproliferation Research and Development (NA-22). The United States Government retains and the publisher, by accepting the article for publication, acknowledges that the United States Government retains a non-exclusive, paid-up, irrevocable, worldwide license to publish or reproduce the published form of this manuscript, or allow others to do so, for United States Government purposes. The U.S. Department of Energy will provide public access to these results of federally sponsored research in accordance with the DOE Public Access Plan (http://energy.gov/downloads/doe-public-access-plan). The views expressed in the article do not necessarily represent the views of the U.S. Department of Energy or the United States Government. DOE/NV/03624--0418.


\bibliographystyle{spiebib}   
\end{document}